\documentclass[aps,twocolumn,showpacs,prl,preprintnumbers,amsmath,amssymb,superscriptaddress]{revtex4-2}
\usepackage{epsf}
\usepackage{graphicx}
\usepackage{sidecap}
 \usepackage{soul}
 \usepackage{array}
 \usepackage{amsmath}
 \usepackage{amssymb}
 \usepackage{dcolumn}
 \usepackage{epstopdf}
 \usepackage{bm}
 \usepackage{amsmath}
 
\usepackage{gensymb}
\usepackage{color}
\usepackage{hyperref}
\usepackage{soul}
\sethlcolor{green}
\usepackage{bbding}
\usepackage{sidecap}
\usepackage{float}

\hypersetup{
    colorlinks=true,
    citecolor=red,
    linkcolor=red,
    filecolor=blue,   
    urlcolor=blue,
}
 
\def \beq {\begin{equation}}
\def \eeq {\end{equation}}
\def\bibsection{\refname}
\renewcommand{\refname}{\noindent\textbf{References}}
\begin{document}
\title{Diverse electronic topography in a distorted kagome metal LaTi$_3$Bi$_4$}
\author{Anup Pradhan Sakhya} \affiliation{Department of Physics, University of Central Florida, Orlando, Florida 32816, USA}  
\author{Brenden R. Ortiz} \affiliation {Materials Science and Technology Division, Oak Ridge National Laboratory, Oak Ridge, Tennessee 37830, USA}
\author{Barun Ghosh} \affiliation{Department of Physics, Northeastern University, Boston, Massachusetts 02115, USA}
\affiliation{Quantum Materials and Sensing Institute, Northeastern University, Burlington, Massachusetts 01803, USA}
\author{Milo Sprague} \affiliation{Department of Physics, University of Central Florida, Orlando, Florida 32816, USA} 
\author{Mazharul Islam Mondal} \affiliation{Department of Physics, University of Central Florida, Orlando, Florida 32816, USA} 
\author{Matthew Matzelle} \affiliation{Department of Physics, Northeastern University, Boston, Massachusetts 02115, USA}
\affiliation{Quantum Materials and Sensing Institute, Northeastern University, Burlington, Massachusetts 01803, USA}
\author{Nabil Atlam} \affiliation{Department of Physics, Northeastern University, Boston, Massachusetts 02115, USA}
\affiliation{Quantum Materials and Sensing Institute, Northeastern University, Burlington, Massachusetts 01803, USA}
\author{Arun K Kumay} \affiliation{Department of Physics, University of Central Florida, Orlando, Florida 32816, USA}
\author{David G. Mandrus} \affiliation{Department of Materials Science and Engineering, The University of Tennessee, Knoxville, Tennessee 37996, USA}
\affiliation {Department of Physics and Astronomy, University of Tennessee Knoxville, Knoxville, Tennessee 37996, USA}
\affiliation{Materials Science and Technology Division, Oak Ridge National Laboratory, Oak Ridge, Tennessee 37831, USA}
\author{Jonathan D. Denlinger} \affiliation{Lawrence Berkeley National Laboratory, Berkeley, CA 94720, USA}
\author{Arun Bansil} \affiliation{Department of Physics, Northeastern University, Boston, Massachusetts 02115, USA}
\affiliation{Quantum Materials and Sensing Institute, Northeastern University, Burlington, Massachusetts 01803, USA}
\author{Madhab Neupane} \thanks{Corresponding author:\href{mailto:madhab.neupane@ucf.edu}{madhab.neupane@ucf.edu}}\affiliation{Department of Physics, University of Central Florida, Orlando, Florida 32816, USA}

\date{\today}

\begin{abstract}
Recent reports on a family of kagome metals of the form $Ln$Ti$_3$Bi$_4$ ($Ln$ = Lanthanide) has stoked interest due to the combination of highly anisotropic magnetism and a rich electronic structure. The electronic structure near the Fermi level is proposed to exhibit Dirac points and van Hove singularities. In this manuscript, we use angle-resolved photoemission spectroscopy measurements in combination with density functional theory calculations to investigate the electronic structure of a newly discovered kagome metal LaTi$_3$Bi$_4$. Our results reveal multiple van Hove singularities (VHSs) with one VHS located in the vicinity of the Fermi level. We clearly observe two flat bands, which originate from the destructive interference of wave functions within the Ti kagome motif. These flat bands and VHSs originate from Ti $d$-orbitals and are very responsive to the polarization of the incident beam. We notice a significant anisotropy in the electronic structure, resulting from the breaking of six-fold rotational symmetry in this material. Our findings demonstrate this new family of Ti based kagome material as a promising platform to explore novel emerging phenomena in the wider $Ln$Ti$_3$Bi$_4$ ($Ln$= lanthanide) family of materials. 
\end{abstract}

\maketitle
\section{I. Introduction}
\indent Kagome motifs owing to their inherent geometric frustration provide an excellent platform to study the interactions between geometry, non-trivial topology, and correlated electronic phenomena \cite{Syozi}. It consists of a hexagon-shaped network of alternating corner-sharing triangles in two dimensions. The band structure of these kagome motifs possess Dirac cones, van Hove singularities (VHSs), and flat bands according to the simple-tight binding model which arises due to its unique lattice geometry \cite{Bergman, Guo, Tang, Kiesel, Norman, Kanoda, Kang, Neupert, Hasan}. Recently, substantial amount of experimental efforts in various magnetic kagome materials have been performed to investigate exotic physics such as giant anomalous Hall effect \cite{Nakatsuji, Nayak, Liu, DChen}, massive Dirac fermions \cite{Ye}, Weyl physics \cite{DFLiu}, intrinsic anomalous Hall conductivity \cite{Ma, Asaba, Zeng, Gao, WMa}, large Berry curvature fields, Dirac fermions and topological Hall effect \cite{Yin2, MLi, Ma, RSLi, Gu, Asaba, Zeng, Y16, Dhakal, Kabir, Lv, QWang}.\\
\indent In addition to these magnetic kagome materials that have been extensively investigated, the discovery of non-magnetic quasi-two dimensional layered kagome superconductors AV$_3$Sb$_5$ (A=K, Rb, and Cs) \cite{Ortiz1, Ortiz2, Wilson} has attracted a lot of interest due to the emergence of Z$_2$ nontrivial band topology \cite {Ortiz2}, charge density wave (CDW) order \cite{Jiang}, intrinsic anomalous Hall effect \cite{Yang135, Yu}, superconductivity and its complex interplay with CDW \cite{Ortiz3, Ortiz4, Yu1, Chen}. 
A related family of Ti-based ATi$_3$Bi$_5$ kagome materials has also been identified, which host nontrivial electronic structure \cite{Seshadri, Yang}, electronic nematicity \cite{YangTi, LiTi}, and 
superconductivity \cite{Yang, GSu}.\\ 
\indent Most recently, another family of kagome metals AM$_3$X$_4$ (A: Lanthanide, Ca, M:V, Ti, X:Sb, Bi) were discovered \cite{Ortiz5, alexander, alexander1, BrendenLn134}. 
The $Ln$M$_3$X$_4$ ($Ln$: Lanthanide) family of materials has generated intense interest lately as they offer a new avenue for studying the interplay between magnetism, spin-orbit coupling, and electronic structure in the kagome motif \cite{alexander, BrendenLn134, anup, Brendentb134, anupce, claudia, hongding}. 
LaTi$_3$Bi$_4$ has also been synthesized and is found to be non-magnetic in nature \cite{BrendenLn134}. However, the experimental electronic structure of this material is not reported yet. 

\begin{figure*} 
\includegraphics[width=16cm]{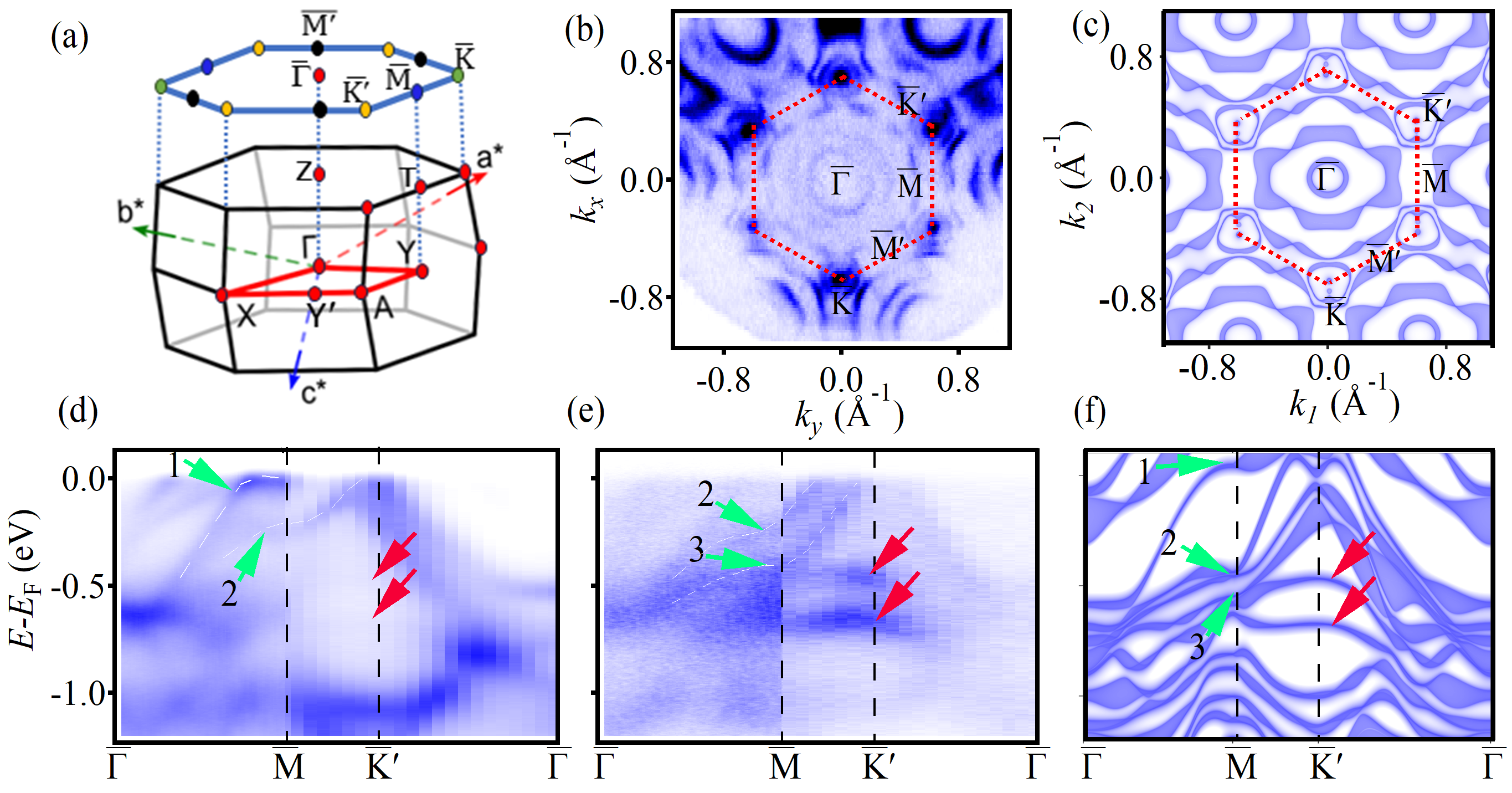} 
   \vspace{-1ex}
	\caption{(a) Bulk three-dimensional Brillouin zone (BZ) along with its projection on the (001) surface. (b) Experimental Fermi surface (FS) measured along the (001) direction using a photon energy of 95 eV and LH polarization. (c) DFT calculated bulk FS. The dotted red hexagon is used to represent the BZ with high-symmetry points as indicated on top of the image. ARPES measured band dispersion using a photon energy of 110 eV along the $\overline{\Gamma}$--$\overline{\text{M}}$--$\overline{\text{K}}'$--$\overline{\Gamma}$ high-symmetry directions with (d) LH polarization and (e) LV polarization. (f) DFT projected bulk band structure along the $\overline{\Gamma}$--$\overline{\text{M}}$--$\overline{\text{K}}'$--$\overline{\Gamma}$ high-symmetry directions. Green arrows indicate VHSs and red arrows indicate the flat bands. ARPES measurements were performed at the ALS beamline 4.0.3 and the SSRL beamline 5-2.}
\label{fig1}
\end{figure*}
\indent In this communiction, we present a detailed exploration of the electronic structure of a new distorted kagome material LaTi$_3$Bi$_4$. Our investigation employs high-resolution ARPES measurements complemented by DFT calculations. LaTi$_3$Bi$_4$ is easily cleavable, possess a layered crystal structure which is ideal for ARPES experiments, allowing us to identify spectroscopic signatures such as multiple VHSs and two flat bands. The distinctive flat bands are a result of the destructive interference of the Ti wavefunction, which forms the kagome motif within the metal. Remarkably, our ARPES results unveil the presence of significant anisotropy in its electronic structure and the presence of linear Dirac-like state at the $\overline{\text{K}}$ high-symmetry points. The VHSs and flat bands display sensitivity to the polarization of incident light, allowing us to extract the orbital character of these features. Our DFT calculations are in good agreement with our experimental results. Consequently, our results establish LaTi$_3$Bi$_4$, as a promising, easily exfoliable novel Ti based kagome metal offering a unique combination of multiple VHSs, two flat bands and anisotropic electronic structure. This opens up exciting opportunities to delve into the intricate interplay among geometry, symmetry, and electron correlation within the broader $Ln$M$_3$X$_4$ family of materials.

\section{II. Methods}
\noindent\textbf{Single crystal growth and characterization:}\\
\indent High-quality single crystals of LaTi$_3$Bi$_4$ were grown through a bismuth self-flux technique which has been described elsewhere in detail \cite{BrendenLn134}.\\ 

\noindent\textbf{ARPES measurements:}\\
\indent ARPES measurements were performed at the Advanced Light Source Beamline 4.0.3 in the Berkeley National Laboratory equipped with Scienta R8000 hemispherical electron analyzer. The angular and energy resolution were set better than 0.2\degree and 15 meV, respectively. All of the sample preparation was performed inside a glove box to avoid oxidation. High-quality flat and shiny crystals of LaTi$_3$Bi$_4$ were cut into small pieces and mounted on copper posts. These crystals were attached to coppers posts using Torr seal. Ceramic posts were then attached on top of the samples. These samples were then loaded into the chamber, after which the chamber was cooled and pumped down for some hours. The crystals were cleaved in-situ at 11 K by knocking off the ceramic posts using a cleaver. The pressure in the UHV chamber was better than 3$\times$10$^{-11}$ Torr. The experiments were performed using both linear horizontal (LH) and linear vertical (LV) polarization. Additional ARPES measurements were performed at Stanford Synchrotron Radiation Lightsource (SSRL), beamline 5-2, Menlo park, CA. Measurements were carried out at a temperature of 8 K. The pressure in the UHV was maintained better than 5$\times$10$^{-11}$ Torr. The angular and energy resolution were set better than 0.2\degree and 15 meV, respectively.\\ 

\noindent\textbf{DFT calculations:}\\
\indent The first-principles electronic structure calculations were carried out within the DFT formalism as implemented in the Vienna ab initio simulation package \cite{Kohn, kresse1996efficient, kresse1999}. The details of the experimental setup and DFT calculations are described in the Supplemental Material \cite{Supp}. We used a plane wave basis set and a projector augmented wave (PAW) pseudopotential as implemented in the VASP package to carry out the \textit{ab initio} calculations \cite{Kohn, kresse1996efficient, kresse1999}. The kinetic energy cutoff to limit the plane wave basis was set to 500 eV. We used a $\Gamma$-centered 8$\times$8$\times$8 $k$-grid to perform the Brillouin zone integration. The strongly constrained and appropriately normed (SCAN) functional was adopted in order to capture the correlation effects \cite{Sun, Sun1}. The crystal structure was optimized by fully relaxing the lattice parameters and the atomic positions until the Hellmann-Feynmann forces on each atom became less than 0.001 eV/\AA.

\section{III. Results and discussion} 
\indent  LaTi$_3$Bi$_4$ crystallizes in the orthorhombic space group F\textit{mmm} (No. 69) with a = 5.89(4) \AA, b = 10.3(4) \AA, c = 24.9(4) \AA, and $\alpha$ = $\beta$ = $\gamma$ = 90\degree \cite{BrendenLn134} (see Supplemental Material Fig. S1 \cite{Supp} for details of the crystal structure). The bulk Brillouin zone (BZ) and its projection on the (001) surface is presented in Fig. 1(a) where the high-symmetry points and directions are labeled. Despite the orthorhombic cell (required by the 2-fold symmetry of the zig-zag chains), both the crystal habit and surface BZ maintain a quasi-hexagonal shape. High-symmetry points are denoted following the quasi 6-fold rotational symmetry, yet a subtle C$_2$ distortion divides the six high-symmetry points $\overline{\text{M}}$ and $\overline{\text{K}}$ of the hexagonal BZ into four new points denoted as $\overline{\text{M}}'$ and $\overline{\text{K}}'$, alongside two distinct surface high-symmetry points $\overline{\text{M}}$ and $\overline{\text{K}}$ along the C$_2$ rotational axes. 

\indent To probe the electronic structure of LaTi$_3$Bi$_4$, we conducted ARPES measurements. In Fig. 1(b), we present the experimental Fermi surface (FS) measured at a temperature of 8 K using a photon energy of 95 eV, revealing the complex fermiology of this material. The FS shows pseudo-hexagonal symmetry which is expected due to the crystallographic structure of the material. Two circular-like pockets at the $\overline{\Gamma}$ point, hexagonal pocket along the $\overline{\Gamma}$--$\overline{\text{M}}$($\overline{\text{M}}'$) direction and triangular pocket with a small central circular pocket is observed at the $\overline{\text{K}}$($\overline{\text{K}}'$) point. The observed experimental FS bears a very close resemblance to the DFT calculated FS presented in Fig. 1(c) (See Supplemental Material Fig. S2 for FS including the surface states \cite{Supp}). 

\indent Polarization-dependent ARPES spectra collected along the $\overline{\Gamma}$--$\overline{\text{M}}$--$\overline{\text{K}}'$--$\overline{\Gamma}$ high-symmetry direction is presented in Fig. 1(d) and Fig. 1(e) using linear horizontal (LH) and linear vertical (LV) polarizations, respectively. From the dispersion around the $\overline{\text{M}}$ point, we observe the presence of three saddle points denoted as VHS1, VHS2, and VHS3. VHS1 and VHS3 are found to be strongly dependent on the polarization of light used. VHS1 is found to be intense when LH polarization light is used and absent when using LV polarization. VHS3 is observed when LV polarization light is used and VHS2 is found to be more intense when measured using LV polarization compared to LH polarization. These VHSs labeled as VHS1, VHS2, and VHS3 are observed at the binding energies of $\sim$ -0.015 eV, $\sim$ -0.2 eV, and $\sim$ -0.4 eV, respectively. We also note the presence of flat bands which are found to lie at the energies of $\sim$ -0.5 eV and $\sim$ -0.7 eV. These flat bands are found to be more intense when LV polarized light is used compared to the LH polarized light. This highlights their intricate multi-orbital character.

\begin{figure} 
	\includegraphics[width=8.5cm]{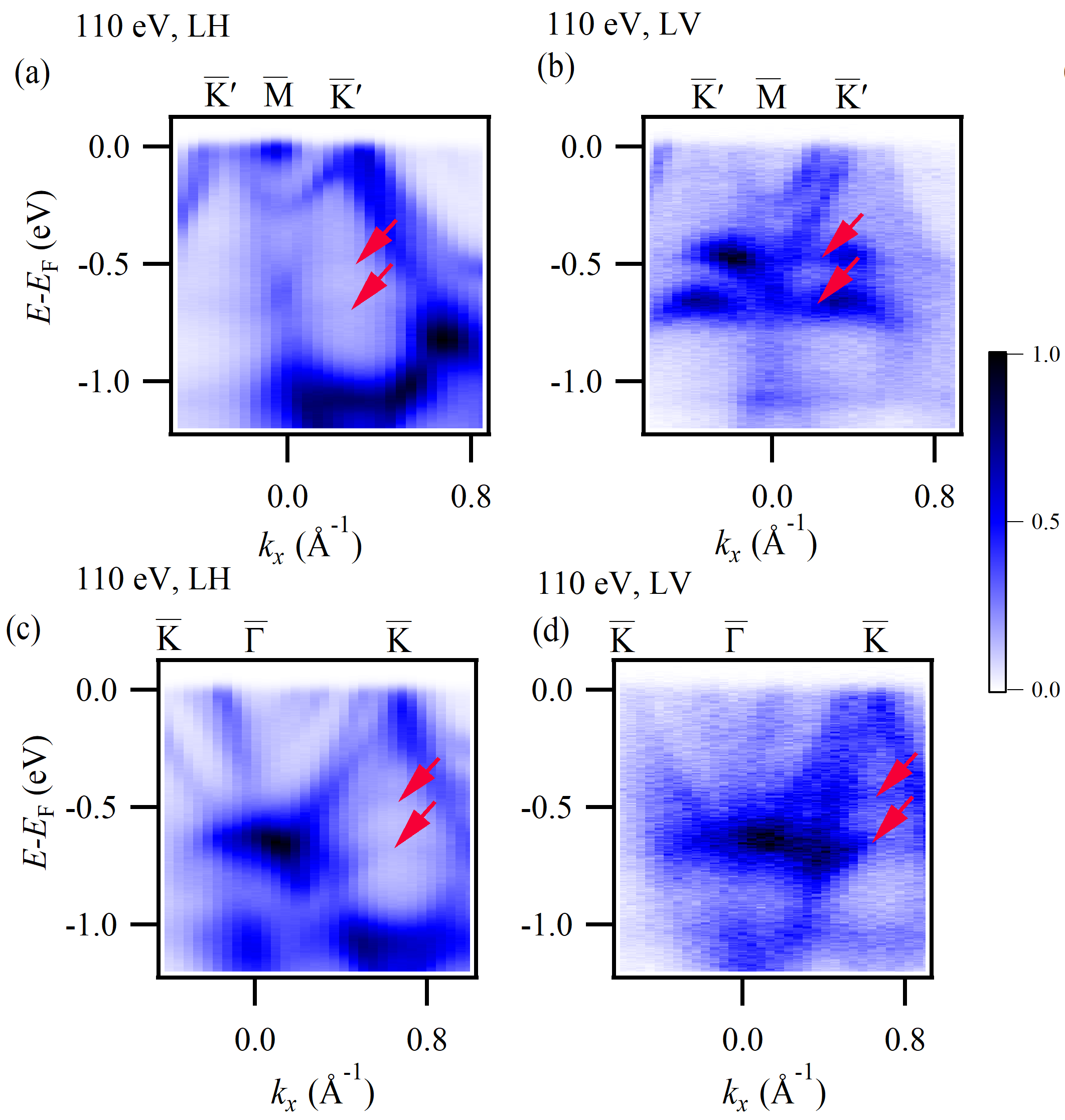} 
    \vspace{-1ex}
	\caption{Observation of flat bands. (a) Experimental band dispersion measured along the $\overline{\text{K}}'$--$\overline{\text{M}}$--$\overline{\text{K}}'$ high-symmetry direction using LH polarization and (b) LV polarization. (c) Experimental band dispersion measured along the $\overline{\text{K}}$--$\overline{\Gamma}$--$\overline{\text{K}}$ high-symmetry direction using LH polarization and (d) LV polarization. 
ARPES measurements were performed at the ALS beamline 4.0.3 using a photon energy of 110 eV and at a temperature of 11 K.}
\label{fig2}
\end{figure}

\indent The band dispersion obtained using polarization dependent ARPES can also be used to gain insights into the sublattice nature of these VHSs from the orbital symmetries. In Fig. S3 of the Supplemental Material \cite{Supp}, we sketch the experimental geometry where the polarization dependent ARPES measurements have been carried out along two orthogonal directions as presented in Fig. 1(d) and Fig. 1(e). According to our experimental geometry (see Supplemental Material Fig. S3 \cite{Supp}), when aligning the $\overline{\Gamma}$--$\overline{\text{M}}$ direction of the sample along the analyzer slit (i.e,. along the \textit{y}-\textit{z} mirror plane), LV polarization suppresses odd orbitals and detects even orbitals. This suggests \textit{$d_{x^{2}-y^{2}}$}, \textit{$d_{z^{2}}$}, and \textit{d$_{yz}$} are detectable in this geometry. However, when aligning the $\overline{\text{K}}'$--$\overline{\text{M}}$ direction of the sample along the analyzer slit (i.e,. along the \textit{x}-\textit{z} mirror plane), LV suppresses even orbitals which indicates \textit{d$_{xy}$} and \textit{d$_{yz}$} are detectable in this geometry. Based on these selection rules and the orbital character of the bands obtained from DFT calculations (see supplemental Material Fig. S4 \cite{Supp}) VHS1 is attributed to \textit{d$_{xy}$} orbital, whereas VHS2, and VHS3 primarily consists of \textit{d$_{yz}$} and \textit{$d_{x^{2}-y^{2}}$},  orbitals, respectively. Using similar arguments, we can also determine the flat bands originates from a mixture of \textit{d$_{xy}$} and \textit{$d_{x^{2}-y^{2}}$} orbitals. The above qualitative arguments, consistent with our DFT calculations, indicates that these three VHSs and flat bands are all attributed to Ti 3\textit{d} orbitals, confirming the crucial role of the 3\textit{d} orbitals of Ti in the observed electronic features. In Fig. 1(f), we present DFT calculated bulk band structure along the $\overline{\Gamma}$--$\overline{\text{M}}$--$\overline{\text{K}}'$--$\overline{\Gamma}$ high-symmetry direction which also reveals the presence of multiple VHSs (see green arrows) and two flat bands (see red arrows) which is in good agreement with our measured ARPES data.\\ 
\indent Polarization-dependent ARPES spectra collected along the $\overline{\text{K}}'$--$\overline{\text{M}}$--$\overline{\text{K}}'$ and $\overline{\text{K}}$--$\overline{\Gamma}$--$\overline{\text{K}}$ high-symmetry directions are also presented in Fig. 2(a-d) (see Supplemental Material Fig. S5 for additional measurement using 90 eV photon energy \cite{Supp}). We observe two hole-like bands at the $\overline{\text{K}}'$ point. Around the $\overline{\Gamma}$ high-symmetry point, two bands are detected which reveal an electron-like curvature. The inner band shows a parabolic dispersion, however, interestingly, the outer band shows a linear dispersion which will be discussed in details in the next section.

\begin{figure}
	\includegraphics[width=8.5cm]{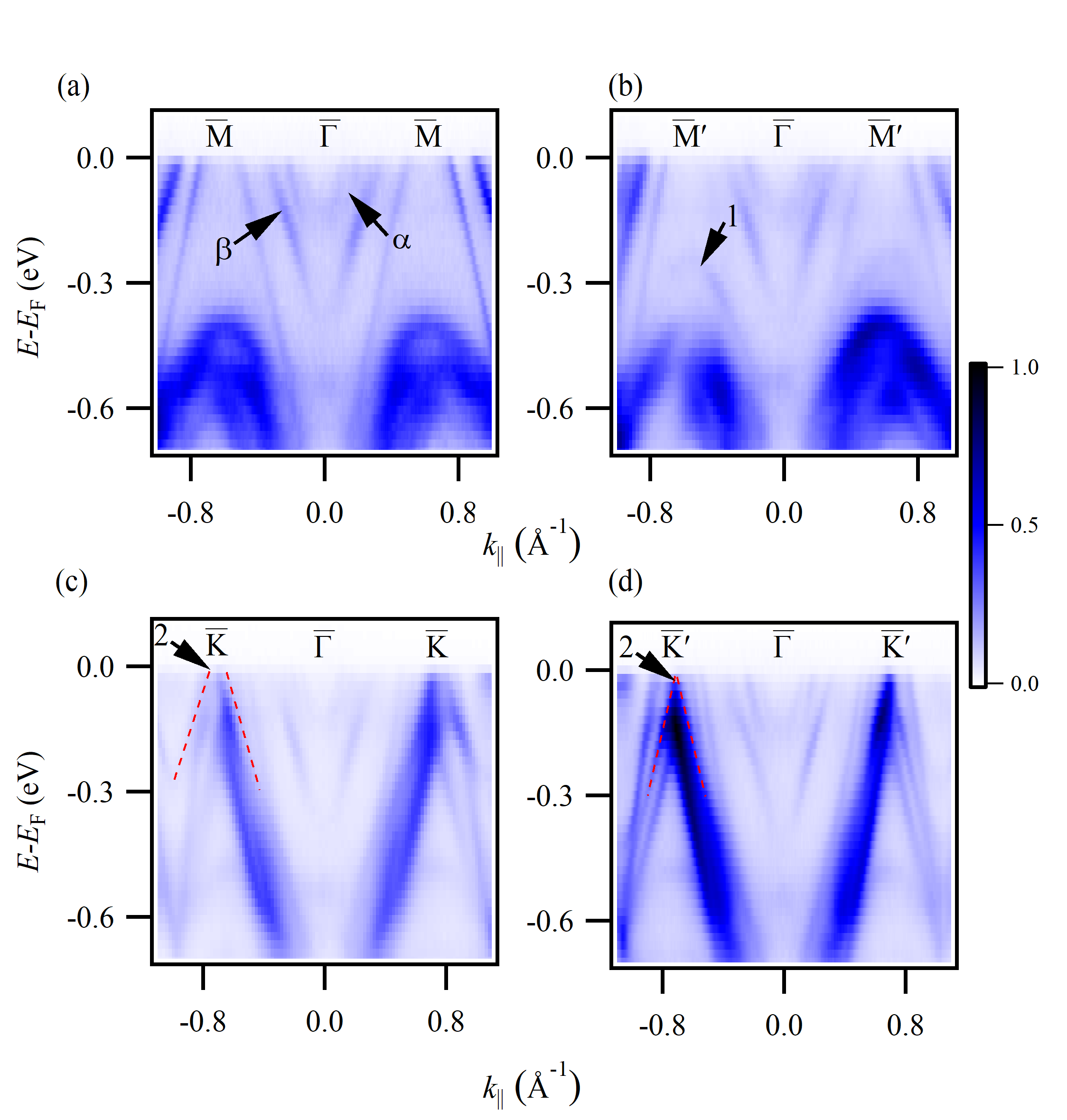} 
    \vspace{-1ex}
	\caption{Anisotropic electronic structure of LaTi$_3$Bi$_4$. (a) Experimental band dispersion measured along the $\overline{\text{M}}$--$\overline{\Gamma}$--$\overline{\text{M}}$ high-symmetry direction using a photon energy of 95 eV. (b) same data as in (a), but measured along the $\overline{\text{M}}'$--$\overline{\Gamma}$--$\overline{\text{M}}'$ high-symmetry direction. (c) Experimental band dispersion measured along the $\overline{\text{K}}$--$\overline{\Gamma}$--$\overline{\text{K}}$ high-symmetry direction using a photon energy of 95 eV. (d) same data as in (c), but measured along the $\overline{\text{K}}'$--$\overline{\Gamma}$--$\overline{\text{K}}'$ high-symmetry direction. ARPES measurements were performed at the SSRL beamline 5-2 at a temperature of 8 K using LH polarization.}
\label{fig3}
\end{figure}

\indent It can be seen that the two sets of bands at the $\overline{\Gamma}$ point are selectively enhanced by LH and LV polarized light, respectively. Both the inner and outer electron-like pockets at the $\overline{\Gamma}$ point are found to be stronger in intensity when using LH polarized light compared to LV polarized light. The selective enhancement of bands by LH and LV polarized light, along with the suppression of intensities, provides additional insights into the orbital characters of these bands. To determine the orbital characters of these bands, we use similar arguments as before. We note LV polarization suppresses the intensity of both the inner parabolic band and the outer linear band, so this band is composed of \textit{d$_{xy}$/\textit{d$_{yz}$}} orbitals. We also note the presence of two flat bands as discussed previously at around $\sim$ -0.5 eV and $\sim$ -0.7 eV, highlighted by the red arrows in Fig. 2(a-d), that extends across a large region of the BZ (see Supplemental Material Fig. S6 for integrated energy distribution curves obtained using 110 eV photon energy \cite{Supp}). The emergence of flat energy bands around $\sim$ -0.5 eV and $\sim$ -0.7 eV can be attributed to destructive interference among wave functions confined within the Ti kagome motif. This interference arises from interlayer interactions between the two kagome layers, leading to the formation of bonding states at lower energies ($\sim$ -0.7 eV) and antibonding states at higher energies ($\sim$ -0.5 eV) \cite{Ye, dispersion}. Through meticulous examination, we observe a slight dispersion within these flat energy bands, which can be explained using a simple tight-binding model. This model demonstrates that distortion of the kagome lattice results in a slightly dispersive flat band and a gapped Dirac cone at the K high-symmetry point. Figure S7 in the Supplemental Material \cite{Supp} compares the electronic band structures obtained from both the distorted and undistorted kagome tight-binding models.

\indent Next, we turn our attention to the effect of structural distortion on the electronic structure of this material. To illustrate this experimentally, we have plotted the band dispersion along the $\overline{\Gamma}$--$\overline{\text{M}}'$--$\overline{\text{K}}$--$\overline{\Gamma}$ high-symmetry lines using both LH and LV polarization, as shown in Supplemental Material Fig. S8 \cite{Supp}. Along this direction, we observed only one VHS at a binding energy of approximately $\sim$ -0.4 eV, which contrasts with the multiple VHSs observed along the $\overline{\Gamma}$--$\overline{\text{M}}$--$\overline{\text{K}}'$--$\overline{\Gamma}$ high-symmetry lines discussed previously in Fig. 1(d,e), highlighting the anisotropy of the electronic structure. Additionally, we have plotted the band dispersion along the $\overline{\text{M}}$--$\overline{\Gamma}$--$\overline{\text{M}}$ and $\overline{\text{M}}'$--$\overline{\Gamma}$--$\overline{\text{M}}'$ high-symmetry lines, as shown in Fig. 3(a-b). Comparing Fig. 3(a) with Fig. 3(b), we observe a significant change in the electronic structure. Specifically, a band labeled as 1 appears around approximately $\sim$ -0.23 eV below the Fermi level in Fig. 3(b), which is not present in Fig. 3(a). A hole-like band is observed at the $\overline{\text{M}}$($\overline{\text{M}}'$) point, while two electron-like bands, labeled as $\alpha$, and $\beta$ encircle the $\overline{\Gamma}$ point, corresponding to the previously discussed circular-like pockets. The innermost band, $\alpha$, displays an electron-like parabolic band with a minimum at approximately \textit{E-E$_F$} = $\sim$ -0.18 eV. This feature is more clearly observed at photon energies ranging from 64 eV to 72 eV in the photon energy-dependent band dispersion plots included in Supplemental Material Fig. S9 \cite{Supp}. However, at higher photon energies, it shows linear behavior, indicating strong dispersion. The band $\beta$, associated with the outer circular-like pocket, extends from the \textit{E$_F$} and merges with the top flat band at approximately \textit{E-E$_F$} = $\sim$ -0.5 eV. Notably, the $\beta$ band exhibits a linear crossing at the $\overline{\Gamma}$ point.
Our bulk and surface projected DFT calculations, included in the Supplemental Material Fig. S10 \cite{Supp}, suggest that the parabolic band $\alpha$ and the linear band $\beta$ at the $\overline{\Gamma}$ high-symmetry point originate from the bulk state. While the DFT-calculated band structure presented in the Supplemental Material Fig. S10 \cite{Supp} agrees well with our ARPES-measured band plots in Fig. 3 and Supplemental Material Fig. S9 \cite{Supp}, the linear band $\beta$ observed in DFT calculations does not extend as far as the experimentally observed $\beta$ band. Such discrepancies between theory and experiment are well-known and may be attributed to effects of the ARPES matrix element \cite{AB1, AB2} and the inherent limitations of DFT in capturing electronic correlations \cite{Sakhyasmbi, anupndsb, anupweyl}.

\indent Next, we examine the band dispersion along the $\overline{\text{K}}$--$\overline{\Gamma}$--$\overline{\text{K}}$ and $\overline{\text{K}}'$--$\overline{\Gamma}$--$\overline{\text{K}}'$ high-symmetry lines, as shown in Fig. 3(c-d). Comparing Fig. 3(c) with Fig. 3(d), we find that the Dirac-like crossing near the $\overline{\text{K}}$ point appears slightly above the Fermi level whereas it lies at the Fermi level for $\overline{\text{K}}'$ high-symmetry point. Ultimately, our findings reveal a pronounced anisotropy in the electronic structure, attributed to the intrinsic structural distortion and breaking of the six-fold rotational symmetry. The material demonstrates a notable divergence from other kagome metals previously studied, as evident from the comparison of band dispersions along various high-symmetry lines. The distorted kagome lattice is responsible for the observed anisotropic electronic structure.

\section{IV. Conclusions}
\indent In summary, our efforts have led to the synthesis of a novel non-magnetic Ti based kagome metal, LaTi$_3$Bi$_4$, which is characterized by its ease of exfoliation. Through a combination of ARPES and DFT calculations, we unveiled multiple VHSs and two flat bands originating from the Ti 3\textit{d} orbitals using a combination of ARPES and DFT calculations. Using polarization-dependent ARPES and DFT calculations, VHS1 is attributed to the \textit{d$_{xy}$} orbital, while VHS2, and VHS3 primarily consist of Ti \textit{d$_{yz}$} and \textit{$d_{x^{2}-y^{2}}$} orbitals. The flat bands originate from a mixture of Ti \textit{d$_{xy}$} and \textit{$d_{x^{2}-y^{2}}$} orbitals. We notice considerable anisotropy in the electronic structure of this material, which arises from the distortion inherent to the crystal structure. 
In a broader perspective, our study opens new avenues for exploring the rich and exotic physics inherent in kagome materials. $Ln$M$_3$X$_4$ (MX: VSb, TiBi) systems emerge as promising domains for investigating the interplay between geometric frustration, magnetism, spin-orbit coupling, and electron correlation, providing an exciting platform for discovering novel physics in these materials.\\ 


\noindent \textbf{ACKNOWLEDGMENTS}\\
\indent Work performed by M.N., M.I.M., and A.K.K. was supported by the US Department of Energy (DOE), Office of Science, Basic Energy Sciences (BES) under Award DE-SC0024304. Work performed by A.P.S., and M.S. was supported by the Air Force Office of Scientific Research MURI, Grant No. FA9550-20-1-0322. Work performed by B.R.O. is sponsored by the Laboratory Directed Research and Development Program of Oak Ridge National Laboratory, managed by UT-Battelle, LLC, for the US Department of Energy. DGM acknowledges support from AFOSR MURI (Novel Light-Matter Interactions in Topologically Non-Trivial Weyl Semimetal Structures and Systems), grant FA9550-20-1-0322. The work at Northeastern University was supported by the Air Force Office of Scientific Research under Award No. FA9550-20-1-0322, and it benefited from the computational resources of Northeastern University’s Advanced Scientific Computation Center (ASCC) and the Discovery Cluster. This research used resources of the Advanced Light Source, a U.S. Department of Energy Office of Science User Facility, under Contract No. DE-AC02-05CH11231.\\

\def\bibsection{\section*{\refname}}

\vspace{2ex}

\end{document}